\pdfoutput = 1
\documentclass[12pt]{scrartcl}
\usepackage[utf8]{inputenc}

\areaset{158mm}{238mm}
\usepackage{setspace}
\usepackage[all]{xypic}
\usepackage{scrlayer-scrpage}

\usepackage{amsmath}
\usepackage{mathtools}
\usepackage{mathrsfs}

\usepackage{amssymb}
\usepackage[unicode=true,linktoc=all]{hyperref}
\usepackage{cite}
\usepackage{bm}
\usepackage{slashed}

\usepackage{tikz,pgf}
\usetikzlibrary{shapes}
\usetikzlibrary{calc}
\usetikzlibrary{decorations.pathmorphing}
\usetikzlibrary{decorations.pathreplacing,shapes.misc}
\usetikzlibrary{positioning}
\usetikzlibrary{decorations.markings}
\tikzset{->-/.style={decoration={
  markings,
  mark=at position .5 with {\arrow{>}}},postaction={decorate}}}
\tikzset{-<-/.style={decoration={
  markings,
  mark=at position .5 with {\arrow{<}}},postaction={decorate}}}

\usepackage{xcolor}
  \definecolor{rblue}{RGB}{81, 49, 193}
  \definecolor{rorange}{RGB}{255, 147, 40}
  \definecolor{rgreen}{RGB}{176, 233, 0}

\newcommand{\fsu}{\mathfrak{su}}
\newcommand{\fso}{\mathfrak{so}}
\newcommand{\fsp}{\mathfrak{sp}}
\newcommand{\fg}{\mathfrak{g}}
\newcommand{\ff}{\mathfrak{f}}
\newcommand{\fe}{\mathfrak{e}}

\newcommand\be{\begin{equation}}
\newcommand\ee{\end{equation}}
\renewcommand{\hat}{\widehat}
\newcommand\MAT[1]{ \left(\begin{matrix} #1 \end{matrix}\right) }

\def\ULST{U(1)^{(1)}_{LST}}
\def\CY{\mathcal{X}}
\def\Fb{\mathcal{S}}
\def\TS{{\mathfrak{P}_T}}

\def\LS{\Delta_0}

\def\TGP{^{2}\mathbf G_{\hat\kappa_\mathscr{P},\hat\kappa_R}}

\DeclareMathOperator{\orb}{orb}

\numberwithin{equation}{section}

\title{2-Group Symmetries of 6d Little String Theories and T-duality}
\author{Michele Del Zotto$^\dagger$ and Kantaro Ohmori$^\sharp$
\\[0.5cm]
	\large\slshape$^\dagger$ Department of Mathematics and Department of Physics and Astronomy, \\[-0.2cm] 
	\large\slshape Uppsala University, Uppsala, Sweden\\
	\large\slshape$^\sharp$ School of Natural Sciences, Institute for Advanced Study,\\[-0.2cm] 
	\large\slshape Princeton, NJ 08540, USA \\[-0.05cm]
	\large\slshape$^\sharp$ Simons Center for Geometry and Physics, SUNY Stony Brook,\\[-0.2cm]
	\large\slshape Stony Brook, NY 11794, USA\\[-0.05cm]
}
\date{}

\begin{document}
\maketitle

\vspace{-13.5cm}
\vspace{14.5cm}
\paragraph{\hspace{.9cm}\large{Abstract}}
\vspace{-.1cm}
\begin{abstract}
\noindent We determine the 2-group structure constants for all the six-dimensional little string theories (LSTs) geometrically engineered in F-theory without frozen singularities. We use this result as a consistency check for T-duality: the 2-groups of a pair of T-dual LSTs have to match. When the T-duality involves a discrete symmetry twist the 2-group used in the matching is modified. We demonstrate the matching of the 2-groups in several examples.
\end{abstract}

\vfill{}
--------------------------

September 2020

\thispagestyle{empty}

\newpage

\tableofcontents

\section{Introduction}

In the past few years we have made lots of progresses about six-dimensional supersymmetric systems decoupled from gravity. Such theories can have either 16 or 8 conserved supercharges, corresponding to $(2, 0)$ and $(1, 1)$, or $(1, 0)$ supersymmetry respectively. In 6D there are two types of UV behaviors: superconformal field theories (SCFTs) and little string theories (LSTs). This was conjectured based upon gauge anomaly cancellation reading between the lines of \cite{Seiberg:1996qx}, and can be argued for using geometric engineering techniques in F-theory \cite{Bhardwaj:2015oru}. The hallmark of these models is given by the presence of strings among their excitations \cite{Strominger:1995ac,Seiberg:1996vs}. SCFTs are characterized by the fact that such strings become all tensionless at the conformal point. LSTs instead have an intrinsic built-in string tension, which entails these systems have T-duality and hence are not ordinary six-dimensional local quantum field theories \cite{Seiberg:1997zk}.\footnote{$\,$ A generalization of QFT that can describe LSTs is proposed in \cite{Kapustin:1999ci}.}
Famous examples of the above are obtained by carefully decoupling gravity from the worldvolume theories of stacks of NS5s respectively in IIA, IIB, and Heterotic superstrings \cite{Seiberg:1997zk,Losev:1997hx}. Some other early foundational works on the subject include \cite{Aspinwall:1996vc,Aspinwall:1997ye,Intriligator:1997dh,Hanany:1997gh,Brunner:1997gf,Intriligator:1999cn,Aharony:1998ub} (see also \cite{Aharony:1999ks} for a review).

\medskip

One of the salient features of LSTs that gives a new window into their interesting exotic dynamics is that they can enjoy a 2-group global symmetry\cite{Cordova:2020tij}.\footnote{$\,$ We refer our readers to the important foundational papers \cite{Baez:2005sn,Sati:2008eg,Sati:2009ic,Fiorenza:2010mh,Fiorenza:2012tb} where the crucial interplay among the Green-Schwarz mechanism and 2-groups (and string 2-Lie algebras) was originally derived.} A 2-group global symmetry \cite{Cordova:2018cvg,Benini:2018reh} is a mixture of a 0-form (i.e.\ ordinary) global symmetry and a 1-form global symmetry (in the language of \cite{Gaiotto:2014kfa}, see also \cite{Sharpe:2015mja}). In particular such 2-group global symmetry in a LST always contains a universal subfactor of the form
\be\label{grupie}
	\TGP = \left(\mathscr P^{(0)} \times SU(2)_R^{(0)}\right) \times_{\hat \kappa_{\mathscr P}, \hat \kappa_R } \ULST,
\ee
where $\mathscr P^{(0)}$ is the 6d Poincar\'e 0-form symmetry group, $SU(2)_R^{(0)}$ is the global $\mathcal N =(1,0)$ R-symmetry of the theory, while $\ULST$ is the 1-form symmetry associated to little string (LS) charge \cite{Cordova:2020tij}. Here $\hat \kappa_{\mathscr P}$ and $\hat \kappa_R$ are the 2-group structure constants, that determine the mixture between higher form global symmetries of different dimensionality. In particular, the invariant background curvature 3-form $H^{(3)}_{LST}$ satisfies a modified Bianchi identity involving the background instanton densities consisting of $\mathscr{P}^{(0)}\times SU(2)_R$ backgrounds and the constants $\hat\kappa_{\mathscr P}$ and $\hat\kappa_R$ (see \eqref{eq:structure}).
These quantities have been computed recently for 6d (1,1) LSTs and for the 6d (1,0) LSTs on NS5 branes of the Heterotic $\text{Spin}(32)/{\mathbb Z}_2$  \cite{Cordova:2020tij}. 

\medskip

The main purpose of this short note is to explore the 2-group symmetries of all the examples of LSTs constructed via F-theory in \cite{Bhardwaj:2015oru}, building upon the classification of 6d SCFTs \cite{Heckman:2013pva,Heckman:2015bfa}.\footnote{ $\,$ See also \cite{Bhardwaj:2015xxa,Tachikawa:2015wka,Bhardwaj:2018jgp}.}. Our results can be summarized in the following

\smallskip

\noindent \textbf{Claims:}  
\begin{itemize}
\item There is a simple formula to compute the structure constants for the 2-group symmetry of all six-dimensional LSTs;\footnote{$\,$ Equation \eqref{eq:FreakOut!}.}
\item The structure constant for the mixture of the one-form symmetry and the $R$-symmetry is always non-zero (for an interacting unitary LST);
\item The 2-groups should be the same between T-dual pairs of LSTs, provided the T-duality does not involve twists;
\item For twisted T-dualities (i.e. T-dualities which involve symmetry twists), a slightly more complicated relation exists.\footnote{$\,$ Notice that it is possible that the T-dual of an untwisted LST is a twisted LST. For example, the T-dual of the 6d $\mathcal{N}=(1,1)$ supersymmetric Yang-Mills LST with a non-simply laced gauge group is a twisted compactification of a 6d $\mathcal{N}=(2,0)$ LST. See sections \ref{sec:LSD} and \ref{chubbychecker}.}
\end{itemize}
We substantiate the claims above by exhibiting several non-trivial examples below. As these examples demonstrate the 2-group structure can provide a useful consistency check for T-dualities among 6d LSTs, which is a useful criterion to exploit in the context of the explosion of 5d dualities among distinct five-dimensional gauge theory phases \cite{Bhardwaj:2019ngx,Apruzzi:2019enx,Bhardwaj:2020gyu} and their applications to the physics of LSTs.\footnote{$\,$ See the IAS seminar \textit{Geometry and 5d N=1 QFT}  by Lakshya Bhardwaj, which is available online at the url:\href{https://video.ias.edu/HET/2020/0330-LakshyaBhardwaj}{https://video.ias.edu/HET/2020/0330-LakshyaBhardwaj} where such applications were announced \cite{prep2}.} The latter was the main motivation for us to publish this study.

\medskip

The structure of this letter is as follows. In section \ref{LSTreview} we give a short coincise review of the geometric engineering of LSTs in F-theory to fix notations and conventions. Along the way, we also begin exploring the geometric engineering counterpart of the field theoretical results of \cite{Cordova:2020tij}, in particular we find the origin of $\ULST$ within the defect group of the corresponding LST. In section \ref{LSTgrupie} the formulas for $\hat \kappa_{\mathscr P}$ and $\hat \kappa_R$ are given. In section \ref{examples} several examples are discussed of T-dual pairs of LSTs both with 16 and 8 supercharges. In all cases we find consistency. In section \ref{LSTKO} we discuss a constraint on endpoints for T-dual pairs which arises from $\hat\kappa_{\mathscr P}$. In section \ref{chubbychecker} the 2-group structure constant matching is generalized to twisted T-dualities and examples with 16 supercharges are demonstrated. 

\medskip

\noindent\textbf{Disclaimer.} To keep this paper short we are not pedagogical: this will benefit the experts, but might make this paper hard to read for a novice. We refer the latter to the first few sections of \cite{DelZotto:2018tcj} or to the review \cite{Heckman:2018jxk} for the necessary background about geometric engineering 6d theories in F-theory, as well as to the paper \cite{Cordova:2020tij} for a beautiful discussion of 2-groups in the context of 6d theories.



\section{LST from F-theory: a lightning review}\label{LSTreview}
In this section to fix notation and conventions we briefly review some aspects of the geometric engineering of (isolated) 6d LSTs in F-theory  \cite{Bhardwaj:2015oru}  that are relevant for our discussion below.\footnote{ $\,$ Throughout this note for simplicity we will assume the backgrounds do not involve $O7^+$ planes, which would require a slightly different formulation of the theory \cite{Bhardwaj:2018jgp}. Our methods can be generalized to that class and it would be interesting to do so.} The F-theory geometric engineering involves a 3-CY which is an elliptic fibration over a two complex-dimensional non-compact K\"ahler surface $\Fb$. The case in which the 3-CY is a general genus-one fibration is relevant for twisted compactifications: the 6d F-theory dynamics is captured by the Jacobian of the genus-one fibration, inequivalent genus-one fibration with the same Jacobian correspond to different twists of the 6d theory down to 5d.\footnote{ $\,$ A Jacobian fibration on the base $\mathcal{S}$ is equivalent to the axio-dilaton background of type IIB string on the base. The choice of a particular 3-CY realizing the given Jacobion corresponds to an additional structure, i.e.\ the twist on $S^1$, in the F-theory/M-theory T-duality \cite{Braun:2014oya,Bhardwaj:2019fzv}.}

The strings of the 6d theory arise from D3 branes wrapping the compact curves in the base, and the lattice $\Lambda \equiv H_2(\Fb,\mathbb Z)$ is identified with the string charge lattice of the theory. The lattice $\Lambda$ is canonically equipped with a quadratic intersection pairing
\be
(\,\cdot\,,\cdot\,) \colon H_2(\Fb,\mathbb Z) \times H_2(\Fb,\mathbb Z) \to \mathbb Z
\ee
which is negative definite in the case of SCFTs and semi-negative definite in the case of LST. Indeed, the former case, by the Artin-Grauert criterion, corresponds to the fact that all the compact curves in the base are shrinkable; in the latter case, on the contrary, the surface $\Fb$ is ruled and has a unique homology class of self intersection zero, which is therefore not shrinkable: its volume defines a scale for this geometry, that is identified with the LST scale. We denote the corresponding curve $\Sigma^0$. Wrapping $\Sigma^0$ with a D3 brane one obtains the BPS little string of the model. Choosing a basis of generators $\Sigma^I$ for $H_2(\Fb,\mathbb Z)$, the negative of the intersection paring
\be\label{eq:DIRAC6D}
\eta^{IJ} \equiv -(\Sigma^I,\Sigma^J)  \qquad I,J = 1,...,r+1
\ee
is identified with the Dirac pairing among BPS strings. The integer $r$ is the rank of the corresponding 6d theory, i.e. the dimension of the corresponding tensor branch. The unit little string charge $\LS \in \Lambda$ is given by the collection of (positive) integers $N_I$ such that
\be
\Sigma^0 = \sum_{I=1}^{r+1} N_I \Sigma^I \qquad \gcd(N_1,...,N_{r+1}) =1 \qquad \LS = (N_1,...,N_{r+1})
\ee
and by construction it corresponds to the (unique) primitive eigenvector of $\eta^{IJ}$ in $\Lambda$ with zero eigenvalue \cite{Bhardwaj:2015oru}.

\medskip

In geometric engineering the vacuum expectation values of the scalar components in the 6d tensormultiplets are identified with the volumes of the curves $\Sigma^I$:
\be
\langle \Phi_I \rangle \sim \text{vol}(\Sigma^I).
\ee
In particular, for a little string theory there is a constraint on the vevs of tensormultiplets scalars
\be\label{eq:LSTscale}
{M_s}^2 = \text{vol}(\Sigma^0) = \sum_{I=1}^{r+1} N_I \, \text{vol}(\Sigma^I) = \sum_{I=1}^{r+1} N_I \, \langle \Phi_I \rangle\,.
\ee
The matrix $\eta^{IJ}$ also encodes the positive semi-definite kinetic matrix for the tensor multiples on the tensor branch of a LST. We denote $b^{(2)}_I$ the dynamical self-dual 2-form fields on the tensor branch. The linear combination
\be\label{eq:background2form}
B^{(2)}_{LST} \equiv \sum_{I=1}^{r+1} N_I \, b^{(2)}_I
\ee
is therefore non-dynamical and it corresponds to the superpartner of the scalar in \eqref{eq:LSTscale}: this is the background 2-form tensor field for the $\ULST$ higher form symmetry.\footnote{$\,$ If the LST is obtained as the worldvolume theory on  NS5 branes decoupled from the gravity sector, this background $\ULST$ is literally the NSNS $B$-field in the decoupled gravity supermultiplet, and the corresponding little string is the fundamental string of the ambient string theory.}

\medskip

\noindent\textbf{Defect group for LSTs.} At this point it is also nice to remark that by an appropriate generalization of the familiar 't Hooft screening argument, all 6d LSTs have a 2-form factor of the defect group of the form  \cite{Bhardwaj:2020phs,Apruzzi:2020zot}
\be
\mathbb D^{(2)} = \mathbb Z \, \oplus \, \bigoplus_{j=1}^p \mathbb Z_{m_j}
\ee
where the integers $m_j>1$ are determined considering the Smith normal form of $\eta^{IJ}$\cite{DelZotto:2015isa,Garcia-Etxebarria:2019cnb,Morrison:2020ool,Albertini:2020mdx,DelZotto:2020esg,Closset:2020scj,Bhardwaj:2020phs,Apruzzi:2020zot}. Notice that the first factor is the term corresponding to the zero eigenvalue giving the LS charge: as remarked in \cite{DelZotto:2020esg} whenever the defect group has a factor $\mathbb Z$ in dimension $d$, we expect to obtain a $U(1)^{(d-1)}$ higher form symmetry.\footnote{$\,$ If we had a $\mathbb{Z}^{(d)}$ symmetry, the background of it would be a $\mathbb{Z}$-valued $d$-cocycle $H^{(d)}$. However it is natural to expect that such a background should actually be realized as the background field strength $H^{(d)} = dB^{(d-1)}$ with a $U(1)^{(d-1)}$ background $B$, in a continuum QFT.}
For the theories we are considering in this paper $d=2$ and we obtain precisely the $\ULST$ higher form symmetry we encountered above.

\medskip

\noindent\textbf{Remark.} In this note we are interested only in the fate of the 2-group structure constants under T-duality. The remaining factors of the defect group are slightly more subtle to analyze because
\begin{itemize}
\item[(a)] upon circle reduction (and twist) factors in $\mathbb D^{(2)}$ can also mix with factors in $\mathbb D^{(1)}$ (associated to the global form of the gauge group of the 6d LST) in non-trivial ways, and 
\item[(b)] if the reduction involves a twist, the corresponding discrete 0-form symmetry can act on the lattice $\Lambda$ and can have a nontrivial interplay with $\mathbb{D}^{(2)}$. 
\end{itemize}
We plan to address these phenomena in more details in future work \cite{prep}.

\section{2-group structure constants and the LS charge}\label{LSTgrupie}

In this section we derive general formulas for the 2-group structure constants $\hat \kappa_{\mathscr{P}}$ and $\hat \kappa_{R}$ and we discuss several applications to untwisted T-dual pairs.

\medskip

Whenever one of the curves in the F-theory base is a part of the discriminant locus of the elliptic fibration, there is a corresponding non-abelian gauge group for which the strings are BPS instantons. In such a case, the corresponding tensor multiplet has a Green-Schwarz coupling, necessary for the cancellation of the gauge anomaly. In facts, all tensor multiplets can be given GS couplings involving background gauge fields for the Poincar\'e symmetry as well as for the other global symmetries of the theory (see e.g. \cite{Sadov:1996zm,Ohmori:2014kda,Intriligator:2014eaa}). As we shall see below, in the case of LSTs, the fact that each dynamical tensor field has such Green-Schwarz coupling generates an interplay among backgrounds for the fields entering in the anomaly polynomial and the 2-form background field for the $\ULST$ higher form symmetry. This is the origin of the 2-group symmetry for LSTs \cite{Cordova:2020tij}.

\medskip



Recall that the dynamical two-form fields $b^{(2)}_I$ have a Green-Schwartz coupling of the form \cite{Ohmori:2014kda}
\begin{equation}
	\eta^{IJ} \int  b^{(2)}_I \wedge \mathbf{X}^{(4)}_J.
\end{equation}
Here the 4-form $\mathbf{X}^{(4)}_J$ can be determined field theoretically for all those tensors that are involved in the cancellation of gauge anomaly. This is always the case for tensors with pairing $\eta^{II}\geq 3$. The only tensors that are not paired to gauge groups in the F-theory construction must have $\eta^{II} = 1$ or $2$. We assume that in the former case the corresponding model is an E-string, in the latter the $\mathcal{N}=(2,0)$ theory of type $\mathfrak{a}_1$. As we have discussed above we are not considering a frozen F-theory geometry. With these assumptions, the GS term $\eta^{IJ}\mathbf{X}^{(4)}_J$ for gravity and R-symmetry background fields is \cite{Ohmori:2014kda,Shimizu:2016lbw,Kim:2016foj} (see also \cite{Cordova:2020tij})
\begin{equation}
\eta^{IJ} \mathbf {X}^{(4)}_J = h^\vee_{\mathfrak{g}_I} c_2(R) + \frac14 (\eta^{II}-2) p_1(TM_6),
\end{equation}
where $h^\vee_{\mathfrak{g}_I}$ is the dual coxeter number of the gauge group $g_I$ coupled with the $I$-th tensormultiplet, and we normalize $h^\vee_{\varnothing}$ to $1$ for the cases $\eta^{II} = 1,2$.\footnote{ $\,$ These two cases in facts are better thought of as having gauge algebras $\mathfrak{sp}_0$ and $\mathfrak{su}_1$ respectively, which indeed have $h^\vee = 1$ \cite{DelZotto:2018tcj} by continuation. We will use both notation interchangeably below.} The index $I$ in $\eta^{II}$ should not be summed.

\medskip

\begin{table}
	\centering
	\label{tab:hvee}
	\begin{tabular}{c|cccccccc}
		$\fg$ & $\fsu(k)$ & $\fso(k)$ & $\fsp(k)$ & $\fe_6$ & $\fe_7$&$\fe_8$& $\fg_2 $ & $\ff_4$\\
		\hline
		$h^\vee$ & $k$ &$k-2$ & $k+1$ & 12 & 18 & 30 & 4 & 9
	\end{tabular}
	\caption{Dual Coxeter numberes}\label{tab:Coxy}
\end{table}

The 2-group structure constants for the LST are captured by the modified Bianchi identity\footnote{ $\,$ We have normalized our characteristic classes with the opposite conventions of \cite{Cordova:2020tij} --- compare our equation \ref{eq:structure} with their (1.18).} for the 2-form background field of $\ULST$ \cite{Cordova:2020tij}
\be\label{eq:structure}
  \frac1{2\pi} d H^{(3)}_{LST} \equiv \hat \kappa_R c_2(R) - \frac{\hat \kappa_\mathscr{P}}{4} p_1(TM_6).
\ee
In presence of GS couplings, all the tensor fields have modified Bianchi identities of the form
\be\label{eq:BIANCHI}
 \frac1{2\pi} d H^{(3)}_{I} = \eta^{IJ} \mathbf {X}^{(4)}_J = h^\vee_{\mathfrak{g}_I} c_2(R) + \frac14 (\eta^{II}-2) p_1(TM_6)
\ee
Now from \eqref{eq:background2form}
\be\label{eq:FLUXUX}
H^{(3)}_{LST} = \sum_{I=1}^{r+1} N_I \, H^{(3)}_{I} 
\ee
Therefore combining \eqref{eq:structure} with \eqref{eq:BIANCHI} and \eqref{eq:FLUXUX} we obtain
\be\label{eq:FreakOut!}
\framebox{$
\phantom{\Bigg|}\hat \kappa_R =   \sum_{I=1}^{r+1} N_I h^\vee_{\mathfrak{g}_I} \qquad \hat \kappa_\mathscr{P} = - \sum_{I=1}^{r+1} N_I (\eta^{II}-2)$}
\ee
thus determining the universal 2-group structure constant for all 6d LSTs from F-theory constructed in \cite{Bhardwaj:2015oru}.

\medskip

\noindent\textbf{Remarks:}
\begin{enumerate}
\item We stress that by including the other global symmetry background gauge fields in $ \eta^{IJ} \mathbf{X}^{(4)}_J$ computing the structure constants for the other factors of the 2-group associated to background fields for the other global symmetries is straightforward by the same method.

\item For the theories with $r=0$ the method here is not strictly speaking applicable, but in the absence of paired tensors the same formula can be derived from the mixed anomaly \cite{Cordova:2020tij}.

\item By anomaly inflow from 6D to 2D \cite{Kim:2016foj,Shimizu:2016lbw}, the modified Bianchi \eqref{eq:BIANCHI} induces the 't Hooft anomaly on the little string. $\hat\kappa_R$ and $\hat\kappa_{\mathscr{P}}$ are the 't Hooft anomalies of $SU(2)_R$ and $SO(4)$ symmetries on the worldsheet. What is special to the little string is that its charge is not gauged, and therefore the little string worldsheet theory and its anomaly in a LST can be directly compared with that of a candidate T-dual LST.
\end{enumerate}




\section{2-groups and T-duality}\label{examples}
Here we see examples of calculations of the structure constants for a few theories, and we apply these as a consistency check for T-dualities: the 2-group structure constants have to match for T-dual pairs of LSTs.
This is simply the natural generalization of symmetry matching we do to check a proposed duality. Combined with the obvious conditions that the 5D rank (i.e.\ the sum of the 6D rank and the ranks of the gauge groups in the tensor branch EFT) should match between a T-dual pair, the 2-group structure constants provides a strict condition for a pair of LSTs to be T-dual.

\subsection{T-duality from geometric engineering}\label{sec:LSD}
Before diving into examples, let us review (and slightly extend) the geometric version of LST T-duality in F-theory \cite{Bhardwaj:2015oru}. We define a pair of 6d LSTs $\mathcal T$ and $\hat{\mathcal T}$ to be \textbf{T-dual} if their (untwisted) circle compactifications give rise to the same 5d KK theory. We define a pair of 6d LSTs $\mathcal T$ and $\hat {\mathcal T}$ to be \textbf{twisted T-dual} if they become equivalent 5d KK theories upon compactification on a circle in which at least one of the two theories is twisted by the action of a (possibly discrete) symmetry. An analogous effect is well-known in the full heterotic string theory \cite{Lerche:1997rr,Witten:1997bs}.  The twisted T-dualities of 6d LSTs are a rather less understood phenomenon. 

\medskip

Let us denote the local 3-CY we are considering $\CY$. As we have reviewed in section \ref{LSTreview} the 6d physics is fully determined by the F-theory of the Jacobian, which we denote $F/\mathcal J_\CY$, where $\mathcal J_\CY$ is an elliptic fibration over the base $\Fb$. We denote the corresponding 6d theory $\mathcal T_{F/\mathcal J_\CY}$. Upon circle compactification we obtain a 5d KK theory $\mathcal T_{M/\CY}$. Geometry seems to suggest there are two cases to be considered
\begin{itemize}
\item \textbf{Case 1}: $\CY$ is elliptically fibered over the base $\Fb$ and thus $\mathcal J_\CY \simeq \CY$: in this case, the theory $\mathcal T_{M/\CY}$ is just the circle reduction of $\mathcal T_{F/\mathcal J_\CY}$ at finite radius by the usual M-theory/F-theory T-duality \cite{Vafa:1996xn};
\item \textbf{Case 2}: $\CY$ is a genus-one fibration over the base $\Fb$: in this case the theory $\mathcal T_{M/\CY}$ is a twisted circle compactification of $\mathcal T_{F/\mathcal J_\CY}$ \cite{Bhardwaj:2019fzv}.
\end{itemize}
\noindent \textbf{LST T-dualities from geometry} \cite{Bhardwaj:2015oru}. The physics of $\mathcal T_{M/\CY}$ is such that the 3-CY $\CY$ and its resolutions correspond to a given chamber on the 5d Coulomb branch.\footnote{$\,$ See e.g. section 3 of \cite{Closset:2018bjz} for a review of the geometric engineering dictionary in M-theory.} Other chambers are realized by flopping $\CY \to \CY^{\boldsymbol{\mu}}$, where we have denoted with $\CY^{\boldsymbol{\mu}}$ the 3-CY obtained from $\CY$ by a sequence of flop transitions $\boldsymbol{\mu}$. If the 3-CY  $\CY^{\boldsymbol{\mu}}$ admits an inequivalent genus-one fibration, over a different base $\hat\Fb$ it will give rise to an inequivalent Jacobian $\hat{\mathcal J}_{\CY^{\boldsymbol{\mu}}}$ and therefore to a different 6d theory ${\mathcal T}_{F/\hat{\mathcal J}_{\CY^{\boldsymbol{\mu}}}}$ obtained from the F-theory of $\hat{\mathcal J}_{\CY^{\boldsymbol{\mu}}}$.

\medskip

\noindent \textbf{Remark.} Notice that in the above discussion $\boldsymbol{\mu}$ could also be the identity (corresponding to no flops): if that is the case $\CY$ itself admits two inequivalent genus-one fibrations.  This is often the case for LSTs of type $\mathcal K$ (in the terminology we introduce in section \ref{LSTKO} below) and many examples studied in \cite{Bhardwaj:2015oru} are of this type --- see also \cite{Anderson:2016cdu}.

\medskip

\noindent Now we can distinguish between the two cases
\begin{itemize}
\item[(a)] \textbf{T-duality}: If $\CY$ and $\CY^{\boldsymbol{\mu}}$ have inequivalent elliptic fibrations, we have a T-duality between the LSTs ${\mathcal T}_{F/{\mathcal J}_{\CY}}$ and ${\mathcal T}_{F/\hat{\mathcal J}_{\CY^{\boldsymbol{\mu}}}}$ 
\item[(b)]  \textbf{Twisted T-duality}:  If $\CY$ and $\CY^{\boldsymbol{\mu}}$ have inequivalent genus-one fibrations of which at least one is not elliptic, we have a twisted T-duality between the LSTs ${\mathcal T}_{F/{\mathcal J}_{\CY}}$ and ${\mathcal T}_{F/\hat{\mathcal J}_{\CY^{\boldsymbol{\mu}}}}$ 
\end{itemize}
Not many examples are known of twisted T-dualities: we discuss some for the case of LSTs with 16 supercharges in section \ref{chubbychecker} below.



\subsection{$\mathcal{N}=(2,0)$ and its T-dual $\mathcal{N}=(1,1)$}\label{sec:16charges}
The $\mathcal N=(2,0)$ LST of type $\mathfrak g \in ADE$ is T-dual to the LST which is the UV completion of the 6d $\mathcal N=(1,1)$ pure SYM gauge theory with simply-laced gauge algebra $\mathfrak{g}$.\footnote{\ There are 6d $\mathcal N=(1,1)$ pure SYM gauge theories with non-simply laced gauge algebras $\mathfrak g \in BCFG$ as well. The T-duals for this class of LSTs are slightly more subtle and are discussed in detail in section \ref{chubbychecker}.} For the $\mathcal N = (2,0)$ case, the string Dirac pairing is identical to the Cartan matrix of the corresponding affine Lie algebra $\mathfrak{g}^{(1)}$. The LS charge coincides with the minimal imaginary root of $\mathfrak{g}^{(1)}$, in other words it is given by
\begin{equation}
	N_I=d_I
\end{equation}
where $d_I$ are the Dynkin (co)marks for the algebra $\mathfrak{g}$ ($d_I$ for the affine nodes is understood to be 1).
The self-intersection number $\eta^{II}$ is $2$ for all $I$.
Therefore, we obtain
\begin{align}
	\hat \kappa_R(\text{type $\mathfrak{g}$ $\mathcal{N}=(2,0)$ LST}) &= \sum_I d_I h^\vee_{\varnothing} = h^\vee_\mathfrak{g}\\
	\hat \kappa_{\mathscr{P}}(\text{type $\mathfrak{g}$ $\mathcal{N}=(2,0)$ LST}) &=0.
\end{align}
On the other hand, from the mixed anomaly argument of \cite{Cordova:2020tij}, the structure constants for the 2-group symmetry of the $\mathcal{N}=(1,1)$ LST of type $\mathfrak{g}$ is 
\begin{align}
	\hat \kappa_R(\text{type $\mathfrak{g}$ $\mathcal{N}=(1,1)$ LST}) &= h^\vee_\mathfrak{g}\\
	\hat \kappa_\mathscr{P}(\text{type $\mathfrak{g}$ $\mathcal{N}=(1,1)$ LST}) &=0.
\end{align}
Clearly we have a match.
Notice that this equality is valid because the 2-group symmetry of the $\mathcal{N}=(2,0)$ LST has a non-trivial structure constant $\kappa_R$ even though the theory does not have any gauge field on its tensor branch. 

\subsection{LST for M5 branes along $S^1 \times \mathbb C^2/\Gamma$ and their T-duals}
As a first $(1,0)$ example we consider slight variaton on the theme in the previous example. Let us consider the LST living on a stack of $K$ M5 branes with transverse space $S^1 \times \mathbb C^2/\Gamma$ in M-theory. The latter is realized in F-theory by a geometry of the form
\be\label{eq:loopG}
\begin{array}{cccccc}
& \mathfrak{g}_\Gamma & \mathfrak{g}_\Gamma & \cdots & \mathfrak{g}_\Gamma & \\
//&2 &2 & \cdots & 2 & //
\end{array}
\ee
where the symbol $//$ indicates that the $K$ curves in the base form a closed loop. The LST which is T-dual to \eqref{eq:loopG} is geometrically engineered with a collection of -2 curves intersecting along an affine $\mathfrak{g}_\Gamma^{(1)}$ diagram, with fiber $I_{d_I \,K}$, where $d_I$ are the corresponding Dynkin labels for each node.
 
\medskip

Whenever $\Gamma \neq \mathbb Z_N$ the above geometry is still singular, and the generalized quiver contains minimal $(\mathfrak{g}_\Gamma,\mathfrak{g}_\Gamma)$ conformal matter \cite{DelZotto:2014hpa} at each collision of -2 curves. For all the above geometries $\hat \kappa_{\mathscr P} = 0$, while $\hat \kappa_R$ is non-trivial. 

\medskip

\noindent\textbf{The case $\Gamma$ is a cyclic subgroup of $SU(2)$.} Let us consider the example $\Gamma= \mathbb Z_N$. In that case the LS charge is $(1,1,1,...,1)$ and the corresponding gauge groups are simply $\mathfrak{su}_N$, therefore 
\be
\hat \kappa_R = KN.
\ee
The symmetry of this formula is not a coincidence, and it is indeed expected: in this case T-duality is precisely swapping the fiber with the base of the fibration in the F-theory geometry
\be
\begin{array}{cccccc}
& \mathfrak{su}_N &  \mathfrak{su}_N & \cdots & \mathfrak{su}_N & \\
//&2_1 &2_2& \cdots & 2_K & //
\end{array} \qquad \overset{T}{\longleftrightarrow} \qquad \begin{array}{cccccc}
& \mathfrak{su}_K &  \mathfrak{su}_K & \cdots & \mathfrak{su}_K & \\
//&2_1 &2_2& \cdots & 2_N & //
\end{array}\,.
\ee
A third T-dual has been proposed in \cite{Bastian:2017ary} for this class of models, that has an F-theory realization
\be
\begin{array}{cccccc}
& \mathfrak{su}_{N K / \ell} &  \mathfrak{su}_{N K / \ell} & \cdots & \mathfrak{su}_{N K / \ell} & \\
//&2_1 &2_2& \cdots & 2_\ell & //
\end{array}
\ee
where $\ell = \text{gcd}(N,K)$. It is straightforward to check that also this model share the same 2-group structure constants, which gives a further consistency check to the proposal of \cite{Bastian:2017ary}.

\medskip

\noindent\textbf{The case $\Gamma$ is a binary dihedral subgroup of $SU(2)$.} If we take $\Gamma$ to be the binary dihedral group of order 8, we have $\mathfrak{g}_\Gamma=\mathfrak{so}_8$, and the corresponding resolved base is
\be
\begin{array}{cccccccccc}
& \mathfrak{so}_8 & \mathfrak{sp}_0 & \mathfrak{so}_8 & \mathfrak{sp}_0 & \cdots & \mathfrak{so}_8 & \mathfrak{sp}_0 & \\
//&4_1 &1_1 &4_2 & 1_2 & \cdots & 4_K & 1_K & //
\end{array}
\ee
where the elementary cell $4,1$ repeats $K$ times. For this theory the LS charge is (1,2,1,2,...,1,2) and we have
\be
\hat \kappa_{\mathscr P } = 0 \qquad\hat \kappa_R = 8 K.
\ee
The corresponding T-dual LST in this case is given by a base that consists of -2 curves intersecting along an affine $\mathfrak{d}_4^{(1)}$ diagram, all supporting a fiber which is of $I_{d_I \, K}$ type where $d_I$ are the Dynkin labels for $\mathfrak{d}_4^{(1)}$. The latter also coincide with the corresponding LS charge $(1,1,1,1,2)$, and therefore
\be
\hat \kappa_R = K + K + K + K + 2 \cdot 2K = 8 K
\ee
as expected. For a general binary dihedral group we obtain $\mathfrak{g}_\Gamma=\mathfrak{so}_{2n}$ and  the corresponding resolved base is
\be\label{eq:dihedral}
\begin{array}{cccccccccc}
& \mathfrak{so}_{2n} & \mathfrak{sp}_{n-4} & \mathfrak{so}_{2n} & \mathfrak{sp}_{n-4}& \cdots & \mathfrak{so}_{2n} & \mathfrak{sp}_{n-4} & \\
//&4_1 &1_1 &4_2 & 1_2 & \cdots & 4_K & 1_K & //
\end{array}\,.
\ee
The LS charge is the same as the structure of the BPS string lattice is unaltered. Therefore we obtain
\be
\hat \kappa_{\mathscr P } = 0 \qquad \qquad \hat \kappa_R = 4 (n-2)K.
\ee
In addition, the 5D rank $r^{5D}$ of the theory is
\be
r^{5D} = 2K -1 + K n + K(n-4) = K(2n-2)-1.
\ee
The T-dual configuration is a collection of -2 curves arranged along an affine $\mathfrak{d}^{(1)}_n$ diagram, with fibers of $I_{d_I \, K}$ type where $d_I$ are the Dynkin labels for $\mathfrak{d}_n^{(1)}$. The latter has rank
\be
r^{5D} = n + 4(K-1) + (n-3)(2K-1) = K(2n-2)-1.
\ee
The LS charge in this case is $(1,1,1,1,2,...,2)$, where $1$'s are for the edge nodes and $2$'s are for the middle nodes, and the structure constants clearly match.
\medskip

\noindent\textbf{Remark}. One might wonder whether there is a third T-dual, as it seems to be the case when $\Gamma$ is cyclic. One can immediately see that there can be no nontrivial T-dualities among the class of theories described by \eqref{eq:dihedral} from the constrains $r^{5D}$ and $\hat\kappa_R$. However, when $K=1$, the $\mathcal{N}=(2,0)$ and $(1,1)$ LSTs of type $\mathfrak{so}(4n-6)$ have the same $r^{5D}$ and $\hat\kappa_R$ as the theory \eqref{eq:dihedral}. It is interesting to determine whether these LSTs are actually T-dual to each other or not.


\medskip

\noindent\textbf{The case $\Gamma$ is an exotic discrete subgroup of $SU(2)$.} We can also consider more exotic examples, for instance for $\Gamma$ the binary tetrahedral discrete subgroup of $SU(2)$, we have $\mathfrak{g}_\Gamma=\mathfrak{e}_6$ and the corresponding resolved base is
\be
\begin{array}{ccccccccccccccccc}
& \mathfrak{sp}_0 & \mathfrak{su}_3 &  \mathfrak{sp}_0 & \mathfrak{e}_6 & \mathfrak{sp}_0 & \mathfrak{su}_3 &  \mathfrak{sp}_0 & \mathfrak{e}_6 &  \cdots &  \mathfrak{sp}_0 & \mathfrak{su}_3 &  \mathfrak{sp}_0 & \mathfrak{e}_6 & \\
//&1_1&3_1&1_1&6_1&1_2&3_2&1_2&6_2& \cdots &1_K&3_K&1_K&6_K& //
\end{array}\,.
\ee
The associated LS charge is $(3,2,3,1\cdots,3,2,3,1)$ and 
\be
\hat \kappa_{\mathscr P } = 0 \qquad\hat \kappa_R = 24 K
\ee
are the 2-group structure constants from our formula. The corresponding T-dual LST in this case is given by a base that consists of -2 curves intersecting along an affine $\mathfrak{e}_6^{(1)}$ diagram, all supporting a fiber which is of $I_{d_I \, K}$ type where $d_I$ are the Dynkin labels for $\mathfrak{e}_6^{(1)}$. The latter also coincide with the corresponding LS charge $(1,1,1,2,2,2,3)$, and therefore
\be
\hat \kappa_R = K + K + K + 2 \cdot 2K + 2 \cdot 2K + 2 \cdot 2K + 3 \cdot 3K = 24K
\ee
as expected.

\subsection{Heterotic instantons probing a $\mathbb C^2 / \mathbb Z_k$ singularity}
Let us consider another simple $\mathcal{N}=(1,0)$ example, which is given by a stack of $N$ Heterotic $E_8\times E_8$ NS5 branes probing the $\mathbb C^2 / \mathbb Z_k$ singularity. The tensor branch geometry is
\footnote{$\,$ The tensor branch of the LST is hard to see in the heterotic string frame, but it becomes evident in the heterotic M-theory frame and its reduction to superstrings of Type I'.}
\begin{equation}
	\begin{array}{ccccccccccccc}
		\varnothing & \varnothing & \fsu(2) & \fsu(3) & \cdots & \fsu(k) & \cdots & \fsu(k) & \fsu(k-1) &\cdots & \fsu(2) & \varnothing &\varnothing\\
		1 & 2 & 2 & 2 & \cdots & 2 & \cdots & 2 & 2 &\cdots & 2 & 2 &1
	\end{array}.
\end{equation}
where the total number of the nodes (including the LST scale) is $N+1$ and (for the sake of notational simplicity) we assume $N+1>2k+1$. The LS charge for this geometry is $(1,1,1,...,1)$. The structure constants are
\begin{align}
	\kappa_R &= (2 + k (k+1) + (N+1-(2k+2))k) = 2 - k^2 + kN\\
	\kappa_\mathscr{p} &= 2.
\end{align}
The dimension of the Coulomb branch after circle compactification to 5d is
\begin{equation}
	r^\text{5d}=\sum_i (\mathrm{rank}(\mathfrak{g}_i)+1) - 1 = 1-k^2+kN.
\end{equation}
The T-duality is the well known $$E_8\times E_8 \quad \overset{T}{\longleftrightarrow} \quad \mathrm{Spin}(32)/\mathbb{Z}_2$$ heterotic T-duality from \cite{Aspinwall:1997ye}. The tensor branch structure of $N$ $\mathrm{Spin}(32)/\mathbb{Z}_2$ instantons probing the $A_{k-1}$ singularity is, when $k$ is even,
\begin{equation}
	\begin{array}{cccccc}
		\fsp(N) & \fsu(2N-8) & \fsu(2N-16) & \cdots & \fsu(2N-4k+8) & \fsp(N-2k) \\
		1 & 2 & 2 & \cdots & 2  &1
	\end{array}.
\end{equation}
as a first check, the 5d rank is
\begin{equation}
	r^\text{5d}= \frac12 \Big(4N-4k+8\Big)\Big(\frac{4k}{8}\Big) + N+N-2k -1 = kN-k^2 +1.
\end{equation}
the structure constants are
\begin{align}
	\kappa_R &= 2 - k^2 + kN\\
	\kappa_\mathscr{p} &= 2. 
\end{align}
however, in those examples, we have only $\fsu$ and $\fsp$ groups which have the property
\begin{equation}
	h^\vee_{\mathfrak {g}} = \mathrm{rank}(\mathfrak{g})+1\,.
\end{equation}
Combined with the fact that the LS charge null vector $N_I =1$ $\forall I$, the matching of $r^\text{5d}$ implies the matching of $\hat\kappa_R$, therefore the structure constant does not give an additional constraint in this case.

\subsection{Heterotic instantons probing an $E_6$ singularity}
As a more non-trivial example, we consider $N$ heterotic instantons probing an $E_6$ (binary tetrahedral) singularity. The tensor branch EFT of both cases are studied in \cite{Aspinwall:1997ye} and expected to be T-dual to each other. The $E_8\times E_8$ side is, for $N=11$, \cite{Aspinwall:1997ye,DelZotto:2014hpa}
\begin{equation}
	\scalebox{0.8}{$\begin{array}{ccccccccccccccccccccccc}
		 &  & \fsu_2 & \fg_2 &  & \ff_4 &  & \fsu_3 &  & \fe_6 &  & \fsu_3 &  &\fe_6 &  & \fsu_3 &  & \ff_4 &  & \fg_2 & \fsu_2 &  &\\
		1 & 2 & 2 & 3 & 1 & 5 & 1 & 3 & 1 & 6 & 1 & 3 & 1 & 6 & 1 & 3 & 1 & 5 & 1 & 3 & 2 & 2 & 1
		\label{eq:e8e8e6}
	\end{array}$}\,.
\end{equation}
Here accounting for $N$ is slightly more complicated due to brane fractionalization. One trick is to find the E-string LST with largest rank into which this theory has a Higgs branch flow. For the case at hand there is a tensor subbranch, where the theory looks like
\begin{equation}
	\begin{array}{ccccccccccccccccccccccc}
		 &  & \fsu_2 & \fg_2 & \ff_4 &  \fe_6 &\fe_6 &  \ff_4 & \fg_2 & \fsu_2 &  &\\
		1 & 2 & 2 & 2 & 2 & 2 & 2 & 2 & 2 & 2 & 2 & 1
	\end{array}
\end{equation}
and from this configuration the theory can flow into the E-string LST with $N=11$. For general $N \ge 10$, the theory should have a tensor subbranch with effective description
\begin{equation}
	\begin{array}{ccccccccccccccccccccccc}
		&  & \fsu_2 & \fg_2 & \ff_4 &  \fe_6 & \cdots & \fe_6 &  \ff_4 & \fg_2 & \fsu_2 &  &\\
		1 & 2 & 2 & 2 & 2 & 2 &\cdots & 2 & 2 & 2 & 2 & 2 & 1
	\end{array}\,,
\end{equation}
and the full tensor branch structure can be obtained by the blow-up method in \cite{Heckman:2013pva}.
The 5d rank is
\begin{equation}
	r^\text{5d}= 12N - 78.
	\label{eq:r5de8e8e6}
\end{equation}
The LS charge for \eqref{eq:e8e8e6} is
\begin{equation}
	 (1, 1, 1, 1, 2, 1, 3, 2, 3, 1, 3, 2, 3, 1, 3, 2, 3, 1, 2, 1, 1, 1, 1).
\end{equation}
For higher $N$, the pattern $1,3,2,3$ in the middle repeats accordingly. We get
\begin{align}
	\hat\kappa_R &= 24 N -166,\\
	\hat\kappa_\mathscr{P} &= 2.
\end{align}
The $\mathrm{Spin}(32)/\mathbb{Z}_2$ side is, from \cite{Aspinwall:1997ye}, 
\begin{equation}
	\begin{array}{ccccc}
		\fsp(N) & \fso(4N-16) & \fsp(3N-24) & \fsu(4N-32) & \fsu(2N-16)\\
		1&4&1&2&2
	\end{array},
	\label{eq:spin32e6}
\end{equation}
when $q\ge8$. The 5d rank of this theory is
\begin{equation}
	r^\text{5d}= 4+ N + 2N-8 +3N-24 + 4N-33 + 2N-17=12 N - 78,
	\label{eq:r5dspin32e6}
\end{equation}
which is the same as \eqref{eq:r5dspin32e6}. The LS charge is (1,1,3,2,1), and the structure constants are
\be
\begin{aligned}
	\hat\kappa_R &= (N+1) +  (4N-18) + 3 (3N-23) + 2(4N-32) + (2N-16)\\&= 24N - 166\\
	\hat\kappa_\mathscr{P} &= 2,
\end{aligned}
\ee
which are consistent with the $E_8\times E_8$ side, as expected.

\section{An endpoint constraint for T-duality from $\hat\kappa_\mathscr{P}$}\label{LSTKO}
Our reader might have noticed that while $\hat\kappa_R$ can take various values, $\hat\kappa_\mathscr{P}$ take only the values $0$ or $2$ in all the above examples. This is actually generally true, at least for the LST constructible in the F-theory without $O7^+$. In \cite{Bhardwaj:2015oru} is found that the endpoint configuration (which is the base after successively shrinking all (-1) curves) for an LST of this kind is either a rational curve with self-intersection number 0, or one of the fibers in the Kodaira classification. Let us call an LST with the former endpoint an LST of type $\mathcal O$, while an LST with the latter endpoint (any out of the Kodaira classification) an LST of type $\mathcal K$.
Since the equation for $\hat\kappa_{\mathscr{P}}$ in \eqref{eq:FreakOut!} has a geometric meaning on the base of F-theory and it is actually an invariant under the blowing-up/down procedure, we can compute the Poincar\'e 2-group structure constant at the endpoint using equation \eqref{eq:FreakOut!} for the endpoint. See appendix \ref{app:proof} for the proof. For all the type $\mathcal O$ LSTs, we have that $\eta(\text{endpoint})=0$ while the LS charge is 1. Therefore,
\begin{equation}
	\hat\kappa_\mathscr{P}(\text{all LST of type $\mathcal O$})= 2.
\end{equation}
On the contrary, for all LSTs of type $\mathcal K$, the endpoints are such that the diagonal components $\eta^{II}(\text{endpoint})$ always equal 2. Therefore we conclude
\begin{equation}
	\hat\kappa_\mathscr{P}(\text{all LST of type $\mathcal K$})= 0.
\end{equation}
An immediate consequence is that the type is a T-duality invariant: an LST of type $\mathcal K$ must be T-dual to another LST of type $\mathcal K$, and an LST of type $\mathcal O$ must be T-dual to another LST of type $\mathcal O$.\footnote{\ It is interesting to observe that for those LSTs that admit a brane engineering within M-theory, the 2-group Poincar\'e structure constant coincides with the number of M9s: $\hat\kappa_\mathscr{P} = \# M9$. This gives an M-theory explanation of the two values we observe for this quantity --- we thank Guglielmo Lockhart for this remark. Moreover, we believe that $\hat\kappa_\mathscr{P} =1$ can be realised for twistings identifying the two M9 branes.}

\medskip

\noindent\textbf{Remark.} The $\mathbb D^{(2)}$ factor of the defect group of this geometry is a blow-up invariant as well \cite{DelZotto:2015isa}. Therefore it depends only on the possible endpoints. We have
\be\label{eq:defecto}
\mathbb D^{(2)} = \begin{cases} \mathbb Z _{LST} & \text{for all LSTs of type }\mathcal O \\ \mathbb Z _{LST} \oplus Z(G_{\mathcal K}) & \text{for all LSTs of type }\mathcal K \end{cases}
\ee
where for a given Kodaira type $\mathcal K$ we denote $G_{\mathcal K}$ the universal cover group corresponding to its split form. For example, if the Kodaira type is $\mathcal I^*_{2n+1}$, we have $G_{\mathcal I^*_{2n+1}} = Spin(2n+9)$ and $\mathbb D^{(2)} =  \mathbb Z _{LST} \oplus \mathbb Z_4$.

\medskip

This prescription has to be modified for LSTs constructed from configurations with $O7^+$ \cite{Bhardwaj:2020phs,Apruzzi:2020zot}, but we will not consider examples of that sort in this note.

\section{Twisted T-dualities and 2-group structure}\label{chubbychecker}
In this section begin a study of the behavior of 2-groups upon twisted T-dualities for 6d LSTs. We focus on LSTs with 16 supercharges as motivating examples. The formalism we develop here extends to the case of LSTs with 8 supercharges.

\subsection{A motivating example}

In Section~\ref{sec:16charges}, we saw that the $\mathcal{N}=(1,1)$ LST of type $\mathfrak{g}$, which is the UV completion of the $\mathcal{N}=(1,1)$ SYM, is T-dual to $\mathcal{N} = (2,0)$ LST of same type, when $\mathfrak{g}$ is one of $A_k$, $D_k$ and $E_{6,7,8}$.
While in the $\mathcal{N}=(1,1)$ side, we can consider a non-simply-laced gauge group, in the $\mathcal{N}=(2,0)$ side the type is restricted to $ADE$.
Therefore we expect the T-duality for a non-simply-laced type should involve a discrete symmetry twist along $S^1$ on $\mathcal{N}=(2,0)$ side.

\medskip

The above expectation is confirmed by the geometrical version of LST T-duality in F-theory along the lines we discussed in section \ref{sec:LSD}. The 6d $(1,1)$ LSTs with non-simply laced gauge groups are of type $\mathcal K$, with an F-theory base that contains a genus one curve, i.e. the $\mathcal I_0$ Kodaira fiber. Since the gauge groups are non-simply laced, the corresponding gauge fibers are \textit{non-split} according to the Tate algorithm \cite{Bershadsky:1996nh} (see also \cite{Katz:2011qp}). Because of the monodromies in the non-split fibers, swapping the $\mathcal I_0$ base with the fiber of the fibration in the non-simply laced case one must obtain a genus-one fibration with nontrivial multisections of order equal to the order of the outer automorphism folding. Therefore geometry predicts we obtain a twisted compactification of a 6d $(2,0)$ LST on $S^1$ where the outer automorphism twist of the corresponding affine Dynkin diagram is acting as a permutation symmetry on the the tensormultiplets, along the lines discussed in section 3.3 of \cite{Bhardwaj:2019fzv}. 

\medskip

To test such a twisted T-duality using the method in this paper requires understanding the behaviour of the 2-group symmetry upon twisting by the action of a discrete symmetry acting on the string charge lattice non-trivially. The full 2-group associated to the resulting 5d KK theory is much more complicated than the mere (continuous) 6d 2-group, even restricting our attention to the 5d KK 2-group corresponding to the universal subgroup $\TGP$ of equation \eqref{grupie} which is the focus of this note. However the latter is mapped to a well-defined subgroup of the 5d KK 2-group: in the following section we compute the structure constants for such subgroup. This requires developing the formalism slightly, to which we now turn.

\subsection{Twisting and 2-group structure constants}

Let us consider a general LST compatified on $S^1$ with twist, and denote the group generated by the twist $\TS$. 
As illustrated in \cite{Bhardwaj:2019fzv,Bhardwaj:2020kim}, the effect of $\TS$ can be read off from the tensor branch EFT, and in general it acts as a combination of permutations of the tensormultiplets and outer automorphisms of the gauge groups and the flavor groups. An interesting effect on the 2-group symmetry occurs when $\TS$ involves a permutation of the tensormultiplets $b^{(2)}_I$. To understand this effect it is necessary to discuss the map of the 6d BPS strings to the BPS strings and particles of the 5d KK theory. 

\medskip

\noindent{\textbf{Mapping 6d BPS strings to the twisted 5d KK theory.}} Let us denote with $[I]$ the $\TS$-orbit of tensor nodes including the node $I$, and with $\orb(\TS)$ the set of such $\TS$-orbits. The 6d BPS strings of the 6d theory give rise to
\begin{itemize}
\item \textbf{BPS strings of the 5d KK theory}. Since only $\TS$-invariant combinations of tensormultiplets survive the twisting, the 6d string charges are mapped to only $|\orb(\TS)|$ integer string charges in the 5d theory. Strings whose charges belong to the same $\TS$-orbit are identified. Label the tensormultiples consistently with our choice of basis for the string charge lattice around \eqref{eq:DIRAC6D} (in such a way that $N_I$ is the charge measured by the $I$-th tensor). A boundstate of 6d BPS strings with charge $$\Delta = (N_1,...,N_{r+1})\,,$$ corresponding to $N_J$ D3 branes wrapping the curve $\Sigma^J$, maps to a boundstate of 5d BPS strings with  charge
\be
P_\TS\Delta =  \left(N^\text{twist}_{[I]}\right)_{[I]\in \orb(\TS)} \qquad N^\text{twist}_{[I]} \equiv \sum_{I \in [I]} N_I\,.
\ee

\item \textbf{BPS particles of the 5d KK theory}. Wrapping a BPS string on the KK $S^1$ gives rise to a 5d BPS particle. In the untwisted case, this establishes an embedding of the 6d string charge lattice into the 5d KK theory particle charge lattice. In the twisted case, instead, not all possible strings can be wrapped on the KK $S^1$ consistently: because of the action of $\TS$ on the strings, only boundstates of 6d BPS strings that are left invariant by the action of $\TS$ can wrap the KK $S^1$ in this case (this is an effect similar to twisted sectors in orbifold CFTs). This implies that the string charges of such states occur in closed orbits of $\TS$ and are therefore also parametrized by $|\orb(\TS)|$ integers. We define the charge of such BPS particles 
\be
\gamma^\TS = \left(Q^\text{wrap}_{[I]}\right)_{[I]\in \orb(\TS)}
\ee
normalized so that the particle coming from the minimal set of strings belonging to $[I]$ has charge $Q^\text{wrap}_{[I]}=1$.
\end{itemize}
In this section we focus on the subsector of the BPS spectrum of the $\TS$-twisted 5d KK theory generated by the two kinds of BPS excitations above. In five-dimensional theories strings and particles can be mutually non-local, and this is indeed the case for the subsector of interest, which follows by

\medskip

\noindent{\textbf{Mapping 6d Dirac pairing to the twisted 5d KK theory and $U(1)^{(1)}$ symmetry.}} The 6d Dirac paring among BPS strings, induces a non-trivial Dirac pairing between the BPS strings and particles we have discussed above: 
\begin{equation}
	\langle \gamma^\TS,P_\TS\Delta\rangle_D  = \sum_{[I], [J]\in \orb(\TS)} Q^\text{wrap}_{[I]} \eta^{[I][J]}_{\TS} N^\text{twist}_{[J]} := \sum_{I,[J]} Q^\text{wrap}_I \eta^{IJ} N^\text{twist}_{[J]}.
\end{equation}
where $Q^\text{wrap}_I$ is the same integer for all $I\in [I]$ by construction.\footnote{$\,$ We have a sum over $I\in[I]$ because the particle of type $[I]$ consists of all the strings of type $I\in[I]$, as explained above.} In other words, the pairing matrix $\eta_{\TS}^{[I][J]}$, whose size is the number of $\TS$-orbits, is 
\begin{equation}
	\eta^{[I][J]}_{\TS} = \sum_{I\in [I]} \eta^{IJ'} \quad\text{for an arbitrary $J'\in[J]$},
\label{eq:etatwist}
\end{equation}
which was also introduced in \cite{Bhardwaj:2019fzv}. Note that \eqref{eq:etatwist} is no longer symmetric as it is a pairing between objects of different dimensionalities.

\medskip

Now we can consider the defect group of the twisted 5d KK theory obtained from $\eta_{\TS}$. In particular, we have a $U(1)^{(1)}$ form symmetry corresponding to the string with the minimal nonzero string charge $\Delta_0^{\TS} = (N_{[I]}^{\TS})_{[I]\in\orb(\TS)}$ satisfying
\begin{equation}
	\sum_{[J]\in \orb({\TS})}\eta^{[I][J]}_{\TS} N_{[J]}^{\TS} = 0,
\end{equation}
that is the primitive right-null-vector of $\eta_{\TS}$.

\medskip

\noindent{\textbf{The 2-group structure constants of the twisted KK theory.}} The 2-group structure constants $\hat\kappa^{\TS}_R$ and $\hat\kappa_{\mathscr P}^{\TS}$ for the twisted compactified theory can be obtained by just replacing the untwisted LS charge $\Delta_0$ by $\Delta_0^{\TS}$ in equation \eqref{eq:FreakOut!}
\begin{equation}
\hat \kappa_R^{\TS} =   \sum_{[I]\in\orb(\TS)} N_{[I]}^{\TS} h^\vee_{\mathfrak{g}_I} \qquad \hat \kappa_\mathscr{P}^{\TS} = - \sum_{[I]\in\orb(\TS)} N_{[I]}^{\TS} (\eta^{II}-2).
\end{equation}

\medskip

\noindent\textbf{Twisting and fractionalization.} Note that the charge of untwisted LS in the compactified theory, $P_\TS \Delta_0$, is also a right-null-vector, and thus it is proportional to $\Delta_0^{\TS}$ up to an integer constant $F$:
\begin{equation}
	P_\TS \Delta_0 = F \Delta_0^{\TS}.
\end{equation}
$F$ being greater than one means that the twisted theory has a fractional little string with fractionality $F$. Correspondingly, the 2-group structure constants for twisted and untwisted cases are also related by the fractionality index $F$:\footnote{$\,$ This fractionalization and the rescaling of the 2-group structure constant should be a consequence of the detailed structure of the symmetry in the 6D LST. If the symmetry in 6D is the direct product of $\TS$ and the continuous 2-group, such a rescale does not happen. Therefore, we expect $\TS$, continuous 2-group, and other discrete (higher-form) symmetries to form a more general higher-group.}
\begin{equation}
	\hat\kappa_{R,\mathscr{P}} = F \, \hat\kappa_{R,\mathscr{P}}^{\TS}.
\end{equation}

If two LSTs $\mathsf{LST}_1$ and $\mathsf{LST}_2$ are dual to each other with twist ${\TS}_1$ and ${\TS}_2$ respectively, the 2-group structure constant should match:
\begin{equation}
	\hat\kappa_{R,\mathscr{P}}^{\mathsf{LST}_1,\TS_1} = \hat\kappa_{R,\mathscr{P}}^{\mathsf{LST}_2,\TS_2}.
	\label{eq:const_twist}
\end{equation}
The fractionality constant $F$ for each twist does not have to match: T-duality does not preserve twisting.

\subsection{Twisted T-duality and non-simply laced 6d $(1,1)$ LSTs}

Coming back to our motivating example. On the gauge theory side the computation of the structure constants is identical, whether the gauge group is simply laced or not. The result is
\begin{align}
	\hat \kappa_R(\text{type $\mathfrak{g}$ $\mathcal{N}=(1,1)$ LST}) &= h^\vee_\mathfrak{g}\\
	\hat \kappa_\mathscr{P}(\text{type $\mathfrak{g}$ $\mathcal{N}=(1,1)$ LST}) &=0.
\end{align}
In the case of the $\mathcal{N}=(2,0)$ LST, the untwisted Dirac paring $\eta$ is the affine Cartan matrix.
The twisting with a permutation symmetry of the affine Dynkin diagram results in $\eta^{\TS}$, which is the (symmetrizable) Cartan matrix for the affine Dynkin diagram obtained by the folding of the original affine diagram.
For the relation between the diagrams and foldings, see e.g.\ \cite{Fuchs:1995zr}.
A natural guess is that the T-dual of the $\mathcal{N}=(2,0)$ LST of type $\mathfrak{g}_A$ with a twist $\TS$ is the $\mathcal{N}=(1,1)$ LST of type $\mathfrak{g}_B$ when the folded affine diagram is that of the untwisted (but not-necessary simply-laced) $\mathfrak{g}_B$ affine algebra.\footnote{$\,$ When the folded affine diagram is of twisted type, we expect the $\mathcal{N}=(1,1)$ would involve an outer automorphism twist on the gauge group. In addition, when the $\mathcal{N}=(1,1)$ side is of $\mathfrak{sp}$ type, there is the discrete theta ambiguity. It would be interesting to study these points.}
In the following we see that this conjuecture is consistent with the constraint \eqref{eq:const_twist} in some expamles.


\medskip

\noindent \textbf{The case of $\mathfrak{g}_B = \mathfrak{g}_2$}. We have that the 6d SYM structure constant is (see table \ref{tab:Coxy})
\be
\hat \kappa_R =  h^{\vee}_{\mathfrak{g}_2} = 4
\ee
We claim that the LST is a twisted T-dual to the (2,0) $\mathfrak{e}_6$ LST. Indeed the collection of curves for the (2,0) $\mathfrak{e}_6$ LST is organized along an affine $\mathfrak{e}_6^{(1)}$ Dynkin diagram
\be
\begin{gathered}
\xymatrix{
&&2_7\ar@{-}[d]\\
&&2_3\\
2_5\ar@{-}[r]&2_2&2_1\ar@{-}[u]\ar@{-}[r]\ar@{-}[l]&2_4\ar@{-}[r]&2_6\\
}
\end{gathered}
\ee
and the folding here is the $\mathbb{Z}_3$ action corresponding to the center $Z(E_6) = \mathbb{Z}_3$ which has orbits
\be
(2_1),(2_2,2_3,2_4),(2_5,2_6,2_7) .
\ee
The corresponding twisted Dirac pairing is
\be
\MAT{2&-1&0\\-3&2&-1\\0&-1&2}
\ee
corresponding to the folding
\be
\mathfrak{e}_6^{(1)} \xrightarrow{\TS^{(3)}} \mathfrak{g}_2^{(1)}\,.
\ee 
The LS charge $\Delta_0 = (3,2,2,2,1,1,1)$ is mapped by such folding to $P_\TS \Delta_0 = (3,6,3)$ and we have fractional little string with charge $\Delta_0^{\TS^{(3)}} = \frac{1}{3}P_\TS \Delta_0 = (1,2,1)$. Since the corresponding $\mathcal{N}=(2,0)$ model does not have any gauge groups, the corresponding structure constant $\hat\kappa_R^{\TS^{(3)}}$ is $(1+2+1) = 4$, which matches with the $\mathcal{N}=(1,1)$ side.

\medskip

\noindent \textbf{The case of $\mathfrak{g}_B = \mathfrak{f}_4$}. We have that the 6d SYM structure constant is (see table \ref{tab:Coxy})
\be
\hat \kappa_R = h^{\vee}_{\mathfrak{f}_4} =  9
\ee
We claim that the latter is a twisted T-dual to the (2,0) $\mathfrak{e}_7$ LST. Indeed the collection of curves for the (2,0) $\mathfrak{e}_7$ LST is organized along an affine $\mathfrak{e}_7^{(1)}$ Dynkin diagram, and the folding here is the outer outomorphism twist
\be
\mathfrak{e}_7^{(1)} \xrightarrow{\TS^{(2)}} \mathfrak{f}_4^{(1)}\,.
\ee 
The LS charge $\Delta_0 = (4,3,3,2,2,2,1,1)$ is mapped by such folding to $(4,6,4,2,2)$, which is divisible by 2. Since the corresponding model does not have any gauge groups, the corresponding structure constant $\hat\kappa_R^{\TS^{(2)}}$ is $(4+6+4+2+2)/2 = 9$, which is consistent with $\mathcal{N}=(1,1)$ side.

\medskip

\noindent\textbf{Remark.} The case of twisted T-duals for 6d LSTs with 8 supercharges is much richer and unexplored.\footnote{\ Some exploratory studies of related backgrounds in F-theory have appeared in \cite{Anderson:2018heq,Anderson:2019kmx,Kimura:2019bzv,Kimura:2019syr}.} The constraint \eqref{eq:const_twist} should be exploited to study the space of twisted T-dual LSTs.

\section*{Acknowledgements}
MDZ thanks Iñaki Garc\'ia Extebarria and Paul-Konstantin Oehlmann for discussions on related topics. KO thanks Clay Córdova and Lakshya Bhardwaj for discussions on the topic.
This project has received funding from the European Research Council
(ERC) under the European Union's Horizon 2020 research and innovation
programme (grant agreement No. 851931). 

\appendix

\section{Proof that $\hat\kappa_\mathscr{P}$ is invariant under blow-ups}\label{app:proof}
Let us relabel the curves from $0$ to $r$ in such a way that the curve to be blown down is the curve number $0$. We will assume to have an LS charge $\Delta_0 = (N_0,N_1,...,N_r)$ with $\text{gcd}(N_0,...,N_r) = 1$. Let us denote $$\ell = \text{gcd}(N_1,...,N_r).$$ We have that $
\eta^{00} = 1$. In the argument below, for the sake of simplicity, we will assume that $\eta^{0J}$ is either 0 or $-1$. For LSTs of sufficiently high rank this is not a restriction, but there are few exceptional cases that violates this assumption in very low ranks. We will comment about them at the end of the argument. With our assumptions, the Dirac pairing for the blown down curve configuration is simply
\be
\hat \eta^{IJ} =   \eta^{IJ} - \eta^{0I}\eta^{0J} \ee
where, slightly abusing notation, the indexes of $\hat \eta^{IJ}$ run from 1 to $r$. The LS charge for $\eta^{IJ}$ is such that $\eta^{IJ}N_J = 0$ which entails that
\be\label{eq:NO}
N_0 = - \sum_{J=1}^r \eta^{0J} N_J
\ee
Therefore
\be
\begin{aligned}
0 = \sum_{J=0}^r \eta^{IJ}N_J &= \eta^{I0}N_0 + \sum_{J=1}^r \eta^{IJ} N_J \\
&= - \eta^{I0} \sum_{J=1}^r \eta^{0J} N_J + \sum_{J=1}^r \eta^{IJ} N_J\\
&= \sum_{J=1}^r  (\eta^{IJ} - \eta^{0I}\eta^{0J})N_J\\
&= \sum_{J=1}^r \hat \eta^{IJ} N_J
\end{aligned}
\ee
We can take $\hat \Delta_0 = (N_1,...,N_r)$ as the LS charge for the blown down configuration provided $\ell = 1$. Assume that that is not the case, then by \eqref{eq:NO} we have that $N_0$ should be divisible by $\ell$ as well, but this is in contraddiction with the fact that $\text{gcd}(N_0,...,N_r) = 1$. 

Now we have that from equation \eqref{eq:FreakOut!}
\be
\begin{aligned}
\hat\kappa_{\mathscr P} &= \sum_{I=0}^r N_I(\eta^{II}-2) =  - N_0 + \sum_{I=1}^r N_I(\eta^{II} - 2)\\
&= \sum_{J=1}^r \eta^{0J} N_J + \sum_{I=1}^r N_I(\eta^{II} - 2) \\
&=- \sum_{J=1}^r \eta^{0J}\eta^{0J} N_J  + \sum_{I=1}^r N_I(\eta^{II} - 2) \qquad ({\text{by our assumption on $\eta^{0J}$}} )\\
&= \sum_{I=1}^r N_I (\eta^{II} -\eta^{0I}\eta^{0I} - 2) = \sum_{I=1}^r N_I (\hat\eta^{II} -2)= \hat\kappa_{\mathscr P}\Big|_\text{blow-down}
\end{aligned}
\ee
Thus establishing that $\hat\kappa_{\mathscr P}  =  \hat\kappa_{\mathscr P}\Big|_\text{blow-down}$ for all LSTs that satisfy our assumptions.
By recursively applying the above, we conclude that $\hat\kappa_{\mathscr P}$ can be computed from the endpoint configuration.

\medskip

Let's consider the exceptional LSTs that are such that $\eta^{00}=1$ and there is an $I$ such that $\eta^{0I}\neq 0,1$. In fact, there is a single such case (without $O7^+$), whose base is
$$\eta = \MAT{4&-2\\-2&1}$$
obtained by blowing up the a Kodaira node $\mathcal I_0$ at the node. The latter is type $\mathcal K$. The uniequness can be understood from the classification of endpoints. It is easy to see that this model has LS charge $\Delta_0={1,2}$ and therefore $\hat\kappa_{\mathscr{P}}=0$. Again it is invariant with respect to blow-down as expected.

\bibliographystyle{ytphys}

\end{document}